\documentstyle[12pt,fleqn,aasms4]{article}

\newcommand{\om}{\mbox{${\Omega}_{0}$ }}
\newcommand{\lam}{\mbox{$\Lambda$ }}
\newcommand{\oml}{\mbox{${\lambda}_{0}$}}
\newcommand{\omlsp}{\mbox{${\lambda}_{0}$ }}
\newcommand{\gam}{\mbox{$\gamma$}}
\newcommand{\al}{\mbox{$\alpha$}}
\newcommand{\be}{\begin{equation}}
\newcommand{\ee}{\end{equation}}
\newcommand{\lab}[1]{\label{#1}}
\newcommand{\r}[1]{~(\ref{#1})}
\newcommand{\p}{\mbox{$\partial$}}

\begin{document}

\title{QUASAR CLUSTERING AND SPACETIME GEOMETRY}

\author{Piotr A. Popowski, David H. Weinberg, Barbara S. Ryden, and Patrick
S. Osmer}
\affil{Ohio State University, Dept. of Astronomy, Columbus, OH 43210}
\affil{E-mail: popowski, dhw, ryden, posmer @astronomy.ohio-state.edu}

\begin{abstract}

The non-Euclidean geometry of spacetime induces an anisotropy in the apparent 
correlation function of high-redshift objects, such as quasars, if redshifts 
and angles are converted to distances
in ``naive'' Euclidean fashion.  The degree of angular distortion
depends on cosmological parameters, especially on the cosmological
constant $\Lambda$, so this effect can constrain $\Lambda$ independent
of any assumptions about the evolution of luminosities, sizes,
or clustering.  We examine the prospects for distinguishing between
low-density ($\Omega_0=0.1-0.4$) cosmological models
with flat and open space geometry using the large quasar samples
anticipated from the Two Degree Field Survey (2dF) and the Sloan Digital
Sky Survey (SDSS).  Along the way, we derive a number of results
that are useful for studies of the quasar correlation function.
In particular, we show that even these large quasar surveys are 
likely to reside in the ``sparse sampling'' regime for correlation
function measurements, so that the statistical fluctuations in measurements
are simply the Poisson fluctuations in the observed numbers of pairs.
As a result: (a) one can devise a simple maximum-likelihood scheme for 
estimating clustering parameters, (b) one can generate Monte Carlo realizations
of correlation function measurements without specifying high-order correlation
functions or creating artificial quasar distributions, and (c) for a fixed
number of quasars, a deeper survey over a smaller area has greater statistical
power than a shallow, large-area survey.  If the quasar correlation length
is equal to the value implied by recent (quite uncertain) estimates, then
the 2dF and SDSS samples can provide clear discrimination between flat and open
geometries for $\om\leq 0.2$ but only marginal discrimination for
$\om = 0.4$.  Clear discrimination is possible for $\om=0.4$ if the true quasar
correlation length is a factor of two larger, and a high-density survey of 
30,000 quasars in 200 square degrees would provide clear discrimination even
for the lower correlation length.  Detection of quasar clustering anisotropy
would confirm the cosmological spacetime curvature that is a fundamental
prediction of general relativity.
\keywords{quasars: general, large-scale structure of Universe}

\end{abstract}

\section{Introduction}

In cosmological models that obey General Relativity and the cosmological
principle, the mass density parameter, $\Omega$, and the cosmological
constant, $\Lambda$, together determine the geometry of spacetime.
The Einstein-de Sitter ($\Omega=1$) model has twin virtues of simplicity:
flat spatial geometry and zero cosmological constant.
However, the ages of globular cluster stars, the high baryon fraction
in galaxy clusters, and some aspects of large scale galaxy clustering
are accounted for more easily in low density models, which have flat 
geometry if 
$\oml \equiv\Lambda c^{2}/3 H_{0}^{2}= 1-\om$\footnote{Subscript 
`zero' indicates the value of the parameters at the present epoch. 
Also, we use the notation 
$H_{0}\equiv 
{\rm h}_{0}\cdot 100\; {\rm km}\; {\rm s}^{-1}\; {\rm Mpc}^{-1}$}, 
or open, negative curvature geometry if $\Lambda=0$.
Dynamical studies of large scale
structure can constrain \om (subject to uncertainties about
biased galaxy formation), but they are insensitive to a cosmological
constant because it represents an unclustered energy component.
This paper examines the prospects for using the anisotropy of the 
quasar correlation function to constrain \om and \oml,
in particular to distinguish between flat and open cosmologies.
While this test is impractical with present quasar samples, the
Anglo-Australian 2-degree Field Survey (hereafter 2dF) and the
Sloan Digital Sky Survey (hereafter SDSS or Sloan) will both produce redshift
samples of several tens of thousands of quasars over the next few years.

Classical methods of measuring spacetime geometry rely on standard candles
or standard rulers, and they are therefore subject to systematic uncertainties
in the evolution of these objects over a large range in redshift.
Alcock \& Paczy\'{n}ski (1979) proposed an alternative approach that is
almost entirely independent of evolution and assumes only that structure
in the universe is statistically isotropic, as implied by the cosmological
principle.  At high redshift, the ratio of distances corresponding to
a given redshift separation $\Delta z$ and angular separation $\Delta \theta$
depends on the spacetime metric, and one can constrain the cosmological
parameters by requiring that they yield isotropic structure.

Alcock \& Paczy\'{n}ski (1979) illustrated their suggestion with the idealized
example of spherical clusters, but nature does not provide us with such
convenient objects of study.  Ryden (1995a) suggested using the statistical
distribution of void shapes in deep galaxy redshift surveys like the SDSS,
but it is not clear that even this million galaxy sample will be sufficient
to detect the effect; it does not extend much beyond $z \sim 0.2$, and
at this depth the geometrical distortion of voids is difficult to separate 
from distortions induced by galaxy peculiar velocities (Ryden \& Melott 1996).
While quasar redshift surveys have fewer tracers, they reach to much
higher redshifts, where the expected geometrical anisotropy is more 
pronounced.  Thus Phillips (1994) suggested using the orientations
of neighboring quasar pairs to implement Alcock \& Paczy\'nski's method.
The 2-point correlation function contains statistical information on
the distribution of all quasar pairs, not just nearest neighbor pairs,
so it is a more powerful measure for detecting distortions of geometry.
It is important to note that these distortions are detectable only because
quasars are clustered.  If quasars were Poisson distributed, then all pair
orientations would be statistically isotropic regardless of the assumed
metric.

Two recent papers (Ballinger, Peacock \& Heavens 1996; 
Matsubara \& Suto 1996; see also Nakamura, Matsubara, \& Suto 1997)
have discussed the use of redshift space clustering in quasar samples
as a tool for constraining \om and \oml.  
Matsubara \& Suto (1996) derive an expression for the correlation
function that includes the effects of geometrical distortion and
linear theory peculiar velocities.  Ballinger et al.\ (1996) derive
expressions for the redshift space power spectrum, including effects of 
geometrical distortion, linear theory peculiar velocities, and incoherent
random velocities generated by small scale collapse.

In the next section, we discuss the geometrical distortion of the
correlation function in the absence of peculiar velocities, extending
the calculation of Phillips (1994).  The distortion due to geometry
alone is simple to understand: changes in the metric just alter the
relation between a $(\Delta z, \Delta \theta)$ separation and the 
corresponding physical distance.  In \S 3 we show that, for realistic
quasar samples, measurement errors in the correlation function are
likely to be dominated by Poisson fluctuations in the number of
quasar pairs.  As a result, we can easily create Monte Carlo realizations
of correlation function measurements for a specified model correlation
function.  In \S 4 we use such Monte Carlo experiments to see what
cosmological constraints can be expected from the 2dF and Sloan quasar
samples.  In \S 5 we summarize the results of these experiments and
discuss some of their limitations.  The most serious of these is
probably our neglect of peculiar velocity distortions in modeling the 
correlation function, but we argue in \S 5 that these are unlikely
to overwhelm the geometrical signal.  A full analysis of observational
data will require joint consideration of velocity and geometry effects,
along the lines envisioned by Ballinger et al.\ (1996).

\section{The redshift space correlation function }
The position of a quasar in 
redshift survey data is characterized by three numbers:
the redshift and two angle coordinates on the sky $(z, \theta_r, \phi_r)$.
For a given quasar, the probability of finding a
neighbor in a given volume element is symmetric about the line of sight,
since this is the only preferred direction.
For any pair
of quasars one can therefore find a transformation 
$(\phi_r, \theta_r) \rightarrow (\phi, \theta) $ 
such that $\phi_1 = \phi_2$ in the new frame of reference.
Consequently, the distance between the points in redshift
space can be described in terms of redshift $z$ and angle $\theta$.


Figure 1 shows the situation with quasars at the positions
$(z_{1}, \theta_{1})$ and $(z_{2}, \theta_{2})$.
We define $s_z \equiv z_{2}-z_{1}$, 
$s_{\theta} \equiv z\cdot (\theta_2- \theta_1)=
z\cdot \Delta\theta $, with $z=(z_{1}+z_{2})/2$ assumed to be much greater
then $z_{2}-z_{1}$. We now adopt the following notation:
\be
s \equiv \sqrt{s_z^2 + s_{\theta}^2}\;, \lab{s.def}
\ee
and
\be
\mu \equiv \frac{s_{z}}{s}\;. \lab{mu.def}
\ee
The angular separation can then be expressed as
\be
{\Delta\theta}^2=z^{-2}s_z^2(\mu^{-2} - 1).  \lab{d.theta}
\ee
In Euclidean geometry, $s_z$ and $s_{\theta}$ would be the line-of-sight
and transverse separations, respectively (in dimensionless, redshift units); 
$s$ would be
the 3-dimensional separation, and $\mu$ would be the cosine of the
angle between the separation vector and the line of sight. 
In reality the geometry is not Euclidean, but we can still keep the formal
definitions of $s$ and $\mu$ introduced above.

Adopting Phillipps' (1994) notation we have
\be
r^2=f^2 {\Delta\theta}^2+g^2 s_z^2\;, \lab{r.sq}
\ee
where in general (e.g., Weinberg 1972)
\be
{\scriptsize
{\normalsize f= \frac{c}{H_{0}}} \times 
\left\{
\begin{array}{ll}
{\displaystyle \frac{1}{\sqrt{1-\om-\oml}} \sinh{\left[\int_{\frac{1}{1+z}}^{1} \frac{\sqrt{1-\om-\oml}dy}{y\sqrt{\frac{\om}{y}-(1-\om-\oml)+\oml y^{2}}}\right]}} & \mbox{if $(\om+\oml)<1$} \\
{\displaystyle \int_{\frac{1}{1+z}}^{1} \frac{dy}{y\sqrt{\frac{\om}{y}+\oml y^{2}}}} & \mbox{if $(\om+\oml)=1$} \\
{\displaystyle \frac{1}{\sqrt{-\left(1-\om-\oml\right)}} \sin{\left[\int_{\frac{1}{1+z}}^{1} \frac{\sqrt{-\left(1-\om-\oml \right)}dy}{y\sqrt{\frac{\om}{y}-\left(1-\om-\oml\right)+\oml y^{2}}}\right]}} & \mbox{if $(\om+\oml)>1$}
\end{array}
\right. 
}\lab{f.def}
\ee
and
\be
g=\frac{c}{H_{0}} \times \frac{1}{\sqrt{\om {(1+z)}^{3} + (1-\om-\oml){(1+z)}^{2}+\oml}}. \lab{g.def}
\ee
Combining equations\r{s.def},\r{mu.def},\r{d.theta} and\r{r.sq}, we find
\be
\frac{r}{s}=g \sqrt{\mu^2 + h^2(1-\mu^2)}\;,\;\;\; \mbox{where} \;\;h\equiv\frac{1}{z}\cdot \frac{f}{g}.  \lab{r.over.s}
\ee

Now suppose that the quasars are clustered and that their
correlation function $\xi(r)$ is described by power-law,
\be
\xi(r)={\left( \frac{r}{r_0} \right)}^{-\gam},  \lab{xi.r}
\ee
where, in principle, the correlation length $r_0$ and the index $\gam$ may
be functions of $z$.
In redshift space, the correlation function at vector separation $(s, \mu)$ 
is simply
the value of $\xi(r)$ at the scalar separation $r$ corresponding to $s$, $\mu$.
Substitution of\r{r.over.s} into\r{xi.r} yields
\be
\xi(s, \mu) = {\left(\frac{s}{s_0}\right)}^{-\gam}{\left[{\mu}^2 + h^2(1-{\mu}^2)\right]}^{-\frac{\gam}{2}}, \lab{xi.s}
\ee
where $s_0=r_0/g$.
The correlation function in these coordinates is anisotropic because the
physical separation corresponding to a given $s$ depends on angle
with respect to the line of sight. The anisotropy is stronger when
$\gam$ is larger because a given change in $r$ then produces a larger change
in $\xi(r)$.

Once $s_0$ and $\gam$ are set, the dependence of 
the correlation function on the geometry of the universe is 
determined entirely through the value of the ``distortion parameter'' $h$.
Figure~2 shows the redshift dependence of $h(z)$
for open and flat cosmologies with various \om.
The functional form of $h(z)$ depends much
more strongly on \omlsp than on \om, implying that clustering anisotropy
can much more easily distinguish models with the same \om and different 
\omlsp than models with similar \omlsp and different 
$\Omega_0$, as Alcock \& Paczy\'{n}ski (1979) pointed out.
There is also an approximate degeneracy in $h(z)$ between open models
with low \om and flat models with relatively high $\Omega_0$.


In the range $1 < z < 5$, $h(z)$ can be well approximated by a 
straight line for most models.  The anisotropy increases with redshift
in all cases, but models can in principle be distinguished both by 
the value of $h$ at a given $z$ and by the overall redshift dependence
of $h(z)$.  As discussed in the following sections, in a real 
redshift survey the information for distinguishing cosmologies
comes mainly from the redshifts where the quasar distribution peaks,
since that is where the clustering can be measured most precisely.

Figure 3 shows the angular dependence of the correlation function,
from equation~(\ref{xi.s}), for various values of $h$.
If $h=1$, the geometry of redshift space is Euclidean, and
there is no correlation function anisotropy.
All models have $h \approx 1$ at $z \ll 1$, and the
$\lambda_0=1$, $\Omega_0=0$ model has $h(z)=1$ at all redshifts.
For other models $h(z)>1$ at high redshifts, and the correlation
function is amplified\footnote{Whether one sees 
``amplification'' or ``suppression'' depends on the choice of reference 
model. Here our reference is Euclidean (\oml=1), but Ballinger et al. (1996)
take \om=1 as a reference and therefore find that alternative
models have suppressed clustering along the line of sight.}
for pair separations along the line of sight ($|\mu|=1$) relative
to pair separations perpendicular to the line of sight.
The anisotropy of the correlation function
increases rapidly with increasing $h$.


\section{Statistical errors in measurements of $\xi(s, \mu)$}
Following Peebles (1980, \S 31), we can define a quasar's
average number of ``clustered neighbors'' by
\be
N_{c}=\int^\infty_0 \bar{n}\xi(r)4\pi r^2dr, \lab{N.c1}
\ee
where $\bar{n}$ is the space density of quasars in a sample.
For purposes of calculation, we assume a power-law correlation function
(equation~[\ref{xi.r}]).
Since the correlation function may fall below this power-law at large
distances, we only consider pairs out to a separation $\alpha r_0$ that
is a small multiple $\alpha$ of the correlation length.
Substitution of\r{xi.r} into\r{N.c1} yields
\be
N_{c}=\frac{4\pi{\al}^{3-\gam}}{3-\gam} \bar{n}r_{0}^3. \lab{N.c2}
\ee
We will take $\al=2$ as a template value in our simulations.
For $\al=2$, the value of $N_c$ from \r{N.c2} varies by only 5\%
for $1 \leq \gam \leq 2$.

Consider a sample of $N_Q$ quasars in an area $A$ square degrees,
and let $F(z)dz$ denote the fraction of
the quasar sample in the redshift range $(z,z+dz)$, normalized
so that $\int_0^\infty F(z) dz=1$.
The observations directly tell us $(N_Q/A)F(z)$, the average
number of quasars per square degree per unit redshift.
The relation between $(N_Q/A)F(z)$
and the number density $\bar{n}(z)$ in
comoving ${\mbox h_0}^{-3}\;\mbox{Mpc}^{3}$ depends on the
cosmological model.
Since the volume of a redshift range $dz$ and solid angle $d\Omega$
is $g\, dz\cdot f^2 d\Omega$,
\be
\bar{n}={3283 (N_Q/A) F(z) \over gf^2}, \lab{bar.n1}
\ee
where 3283 is the number of square degrees per radian.
For compactness, in this and future equations we generally omit the
explicit $z$-dependence of $\bar{n}$, $f$, and $g$.
From\r{f.def} and\r{g.def} it is clear that both $f$ and $g$ involve
the factor $cH_{0}^{-1}=3000 {\mbox h_0}^{-1}\mbox{Mpc}$, and it is convenient
to write them in the form
\be
f=\frac{c}{H_{0}}\tilde{f}, \qquad\mbox{and}\qquad 
g=\frac{c}{H_{0}}\tilde{g}. \lab{f.g.def}
\ee
We use the notation\r{f.g.def} 
and the value of $cH_{0}^{-1}$ to write\r{bar.n1} as
\be
\bar{n}=1.21\cdot 10^{-7} \frac{N_Q}{A} F(z)
{\tilde{g}}^{-1}\tilde{f}^{-2} ~~
{\rm h}_0^3\;{\rm Mpc}^{-3} ~{\rm (comoving)}.
\lab{bar.n2}
\ee
Substitution of\r{bar.n2} into\r{N.c2} gives
\be
N_{c}(z)=\frac{4\pi{\al}^{3-\gam}}{3-\gam} r_{0}^3 
\left(1.21\cdot 10^{-7} \frac{N_{Q}}{A} F(z) 
{\tilde{g}}^{-1}\tilde{f}^{-2}\right),\lab{N.c3}
\ee
for $r_0$ in comoving h$_0^{-1}$ Mpc.
The number of correlated pairs in the range $(z, z+dz)$ is
\be
N_{pairs}(z)=N_{Q} F(z) N_{c}(z)\propto \frac{N_{Q}^2}{A} F^2(z) r_{0}^3. \lab{N.pairs1}
\ee

As a fiducial case let us take $\al=2$, $\gam=2$, 
$N_{Q}= 10^{5}$, $A= 10^{4} {\Box}^\circ $, and $F(z)=0.5$.
Then
\be
N_{c}(z)=0.11 {\left( \frac{r_{0}}{10 {\mbox h_0}^{-1}\mbox{Mpc}} \right)}^{3} \left(\frac{{\tilde{g}}^{-1}\tilde{f}^{-2}}{7.27}\right), \lab{N.c4}
\ee
where $r_0$ is the comoving correlation length.
We have scaled ${\tilde{g}}^{-1}\tilde{f}^{-2}$ to the value for an
\om=1 model at $z=2$.
Substitution of\r{N.c4} into\r{N.pairs1} yields
\be
N_{pairs}(z)=5500 {\left( \frac{r_{0}}{10 {\mbox h_0}^{-1}\mbox{Mpc}} \right)}^{3} \left(\frac{{\tilde{g}}^{-1}\tilde{f}^{-2}}{7.27}\right) \lab{N.pairs2}
\ee
for the number of correlated quasar pairs expected in an interval
$\Delta z \approx 1$ near the peak of the redshift distribution $F(z)$
in a large quasar redshift survey.

We now come to the central issue of this Section --- a key issue for
the entire paper, in fact --- the nature of statistical fluctuations in
estimates of the correlation function.  In the limit $N_c \gg 1$,
the statistical uncertainty is dominated by the finite number of
independent structures in the survey volume, an effect sometimes referred to
as ``cosmic variance.''  As a conceptual toy model, one can imagine
the structure to consist of clusters with an average of $N_c$ members
apiece.  Each cluster is sampled by many correlated pairs, so the
fluctuation in the number of clusters within the survey dominates
the error in $\xi$.  In the opposite limit, $N_c \ll 1$, most clusters
are not ``detected'' with even a single correlated pair, and it
is the Poisson fluctuation in the number of pairs that dominates the 
statistical error rather than the fluctuation in the number of
independent structures.  In this ``sparse sampling'' or ``Poisson limit,''
the error in an estimate of $\xi(s,\mu)$ in a bin $(\Delta s, \Delta \mu)$
is simply the Poisson error in the number of pairs in the bin,
and the errors in separate bins are statistically independent.
The transition between the dense sampling and sparse sampling limits
occurs at $N_c \approx 1$; we have carried out numerical experiments
with Monte Carlo clustering models (Soneira \& Peebles 1978) to
confirm that the Poisson error approximation holds quite accurately
for $N_c \la 1$.

In a typical galaxy redshift survey, $N_c(z) \gg 1$ nearby, 
and $N_c(z)$ drops below one at large distances, where only the
most luminous galaxies lie above the survey's apparent magnitude limit.
Roughly speaking, the volume within which $N_c(z) \ga 1$ is
the effective volume of the survey for purposes of estimating $\xi$;
there is some usable information from larger distances, but as
the structure ``fades out'' under increasingly sparse sampling, the
statistical significance of this additional information drops.

Equation~(\ref{N.c4}) implies that even an ambitious quasar
redshift survey is likely to be in the sparse sampling regime
at {\it all} redshifts, unless the comoving quasar correlation
length is substantially larger than $10 {\rm h}_0^{-1}\;$Mpc.
This result has several important implications.
First, the independent, Poisson errors in $\xi$ allow a straightforward
maximum-likelihood scheme for estimating correlation function parameters.
We describe this scheme in \S 4.3 below; this method was developed
independently by Croft et al. (1997), who applied it to rich galaxy
clusters, and it was applied to a sample of high-redshift quasars
from the Palomar Transit Grism Survey by Stephens et al. (1997).
A second implication is that, given a theoretical prediction of
$\xi(s,\mu)$, one can generate Monte Carlo realizations of measured
correlation functions for a given sample without creating detailed
realizations of a clustering pattern, because the fluctuation in
a measured value of $\xi$ depends only on the expected number of pairs
and is independent of fluctuations in other bins and of the higher-order
clustering properties of the quasar distribution.
We describe this method for creating Monte Carlo realizations in \S 4.2
below, and in \S 4.4 we use it to evaluate the power of the SDSS
and 2dF quasar redshift surveys for constraining the cosmological constant.
Finally, there is an observational implication: for a fixed number of 
quasars, a deeper survey over a smaller area has greater statistical
power, at least for measurements of the correlation function on
scales $\sim r_0$.  

This last point is demonstrated most clearly by generalizing
equations~(\ref{N.c4}) and~(\ref{N.pairs2}) to
\be
N_{c}(z)=0.11 \left(\frac{\al^{3-\gam}/(3-\gam)}{2}\right)
{\left(\frac{r_{0}(z)}{10 h^{-1} \mbox{Mpc}}\right)}^3 
\left(\frac{N_{Q}/A}{10/{\Box}^{o}}\right) 
\left(\frac{F(z)}{0.5}\right)
\left(\frac{{\tilde{g}}^{-1}\tilde{f}^{-2}}{7.27}\right),\lab{N.c5}
\ee
\be
N_{pairs}(z)=5500 \left(\frac{\al^{3-\gam}/(3-\gam)}{2}\right)
{\left(\frac{r_{0}(z)}{10 h^{-1} \mbox{Mpc}}\right)}^3 
\left(\frac{N_{Q}}{10^5}\right)
\left(\frac{N_{Q}/A}{10/{\Box}^{o}}\right) 
{\left(\frac{F(z)}{0.5}\right)}^2 
\left(\frac{{\tilde{g}}^{-1}\tilde{f}^{-2}}{7.27}\right). \lab{N.pairs3}
\ee
If $N_c(z) \la 1$, then the effective signal for correlation function
measurements is set by $N_{pairs}(z)$, since
each correlated pair contributes non-redundant information.
At fixed surface density
$N_Q/A$, this signal increases in proportion to the number of
quasars $N_Q$, but at fixed area $A$ it is proportional to $N_Q^2$.
Of course, increasing $N_Q$ at fixed $A$ usually requires one to
observe fainter quasars, while increasing $N_Q$ at fixed surface density 
does not.  Equation~(\ref{N.pairs3}) also implies that the precision of
correlation function measurements will be peaked sharply near the
maximum of the survey's redshift distribution $F(z)$ and that the
precision attained will depend strongly on the actual value
of the quasar correlation length $r_0$.

The remaining equations in this Section are not fundamental,
but they are needed for our Monte Carlo simulations and 
maximum-likelihood parameter estimation, and we include their
derivation for the convenience of others who may wish to 
pursue similar approaches.
A physical model specifies the correlation function in
real space ($r$-space), but a redshift survey provides data
in redshift space ($s$-space).
The relation between coordinates in these two complementary descriptions is
\be
\left( \begin{array}{c}
dr \\ d(\cos t)
\end{array} \right) = \underbrace{\left( \begin{array}{cc} 
\frac{\p r}{\p s} & \frac{\p r}{\p \mu} \\
\frac{\p (\cos t)}{\p s} & \frac{\p (\cos t)}{\p \mu}
\end{array} \right)}_{\mbox{\boldmath $A$}} \left( \begin{array}{c}
ds \\ d \mu
\end{array} \right), \lab{matrix}
\ee
where $\cos{t} = g s_{z}/r$.
Thus
\be
dr \,d(\cos t)= \det{\!\mbox{\boldmath $A$}} \; ds \, d\mu .  \lab{determinant}
\ee
A short calculation leads to
\be
d(\cos{t})=\frac{h^2 d\mu}{[h^2-(h^2-1)\mu^2]^\frac{3}{2}}, \lab{d.cos}
\ee
and
\be
dr=g \sqrt{h^2 - (h^2-1) \mu^2} \,ds - \frac{g s \mu (h^2-1) d\mu}{\sqrt{h^2 - (h^2-1) \mu^2}}. \lab{d.r}
\ee
Since $\frac{\p (\cos t)}{\p s}=0$, equation\r{determinant} yields
\be
(dV)_r = -2\pi r^2 dr d(\cos{t}) = -2\pi r^2 \left[\frac{\p r}{\p s} \cdot \frac{\p (\cos t)}{\p \mu}\right] ds d\mu \lab{vol.1}
\ee
for the volume element in real space.
Substitution of\r{d.cos} \&\r{d.r} into\r{vol.1} leads to
\be
(dV)_r=g^3 h^2 (-2\pi s^2 ds d\mu)= g^3 h^2 (dV)_s, \lab{vol.2}
\ee
where $(dV)_s$ is the volume element in redshift space.
The number of quasars in a given region will be the same
regardless of the adopted coordinate system.  Thus
\be
n_s (dV)_s \equiv n(z) (dV)_r, \lab{vol.3}
\ee
where $n_s$ and $n(z)$ are number densities of quasars in redshift and real
spaces, respectively.
Combining\r{vol.2} and\r{vol.3} leads to 
\be
n_s(s, \mu) = g^3(z) h^2(z) n(z) . \lab{num.den} 
\ee

The number of pairs expected in an
infinitesimally small bin in $(s, \mu)$ within the finite redshift range
$(z, z+\Delta z)$ is
\be
dN_{pairs}=N(z, z+\Delta z) n_s [1+\xi(s, \mu)] (-2\pi s^2 ds d\mu),
\lab{pair.num.1}
\ee
where $N(z, z+\Delta z)$ is the number of observed quasars 
in this redshift range.
Since $F(z)dz$ is the fraction of quasar redshifts in an infinitesimal 
range $dz$, 
\be
N(z, z+\Delta z)=N_{Q} \int_{z}^{z+\Delta z} F(z) dz,
\lab{N.z.dz}
\ee
where $N_Q$ is the total number of quasars in the sample.
Substitution of\r{xi.s},\r{num.den} and\r{N.z.dz} into\r{pair.num.1} for
infinitesimally small $\Delta z\equiv dz$ yields 
\be
dN_{pairs} = -2\pi N_Q g^3 h^2 \bar{n} F(z) 
\left\{1+{\left(\frac{s}{s_0}\right)}^{-\gam}
{\left[{\mu}^2 + h^2(1-{\mu}^2)\right]}^{-\frac{\gam}{2}}\right\} 
s^2 dz ds d\mu .
\lab{dN.pairs}
\ee
Here $s_0 \equiv r_0/g$ is the line-of-sight separation that corresponds
to the comoving correlation length $r_0$, and the 
mean number density $\bar{n}$ is given in terms of
$N_Q$, $A$, and $F(z)$ by equation\r{bar.n2}.
The quantities $g$, $h$, and $\bar{n}$ all depend on 
redshift, and $s_0$ and $\gamma$ may depend on redshift also.

\section{Monte Carlo experiments}

\subsection{Observational perspective}

Successful measurement of quasar clustering anisotropy will
require redshift samples much larger than those that exist today.
There are two clear prospects for such samples, the SDSS and
a quasar redshift survey using the 2dF fiber spectrograph on the 
Anglo-Australian Telescope.
The SDSS plans to obtain redshifts for 80,000 quasars in an area
of 5,000 square degrees in the North Galactic Cap.
It will also measure redshifts for $\sim 20,000$ quasars in the surrounding
5,000 square degree ``skirt,'' but this subset will be less useful for
clustering measurements because of its lower density.
The SDSS will also conduct a deeper quasar survey in a stripe of $\sim 200$
square degrees in the South Galactic Cap, to a limiting magnitude
yet to be determined.
(A discussion of the planned SDSS quasar survey can be found at
http://www.astro.princeton.edu/BBOOK/SCIENCE/QUASARS/quasars.html.)
The SDSS will select quasar candidates based on 5-band CCD photometry.
The planned 2dF survey will target $\sim 30,000$ quasars in an
area $\sim 750$ square degrees, selected as UV-excess stellar objects
on scanned photographic plates (Shanks, private communication).

In order to generate Monte Carlo realizations of correlation function
measurements from these samples, we need to know (approximately) the
expected redshift distributions $F(z)$.
We have computed these using the observational determinations of
the quasar luminosity function by
Boyle (1991) and Warren, Hewett, \& Osmer (1994).
The Boyle (1991) results are based on the Boyle, Shanks \& Peterson (1988) 
survey of quasars selected by the UV-excess method, 
which is effective mainly for $z \la 2.2$.
The Warren et al. (1994)
results are based on a multicolor survey and probe the redshift range
$2.0 < z < 4.5$.  For our calculations, we assume that the 2dF survey
will contain quasars only up to $z=2.2$, because of its UV-excess
selection technique, and that the SDSS multi-color selection will
identify quasars at all redshifts, with apparent magnitude being
the only important limit.  We find that the planned SDSS and 2dF
surface densities imply limiting apparent magnitudes of roughly
$B=19.6$ and $B=21.4$, respectively.  With these apparent magnitude
limits, we compute the number of quasars per square degree per unit
redshift from the above-mentioned luminosity functions.

The solid curves in the two panels of Figure~4 show the results
of these calculations.  Sharp dips at $z \sim 2- 2.2$ are an artifact
of using the Boyle (1991) luminosity function close to the redshift
limit of the Boyle et al. (1988) survey.  More broadly, the shapes
of these curves reflect the interplay between the peak in the quasar 
luminosity
function at $z \sim 2-3$ and the smaller fraction of the luminosity
function that is visible above the apparent magnitude limit at larger
distances.  Of course, $F(z)$ is also affected by the increase of the
differential volume element $dV/dz$ with $z$, especially at low redshift.
Up to $z\sim 2$, the evolution effects and the magnitude limit
work in different directions, producing a relatively flat redshift 
distribution.  Above $z\sim 2$ both the
number density of quasars and our ability to see them decreases, and 
the redshift distribution drops rapidly towards zero.


Since these calculated redshift distributions are only approximate ---
the quasar luminosity function is itself uncertain, and we have not 
modeled the selection criteria and corresponding incompleteness of 
the two surveys in any detail --- we fit them with simple analytic
forms to use in our Monte Carlo calculations below.  These fits
are indicated by the dashed curves in Figure~4.
The survey's target selection algorithms will probably focus on objects
with point-source morphology in order to reduce contamination by galaxies.
This selection technique may exclude quasars at low redshift, where
the host galaxies are bright enough to be detected as extended emission.
In our $F(z)$ fits, we model this effect by a sharp cutoff at $z<0.4$.

\subsection{Generating Monte Carlo simulations}
For our Monte Carlo experiments, we wish to generate realizations
of {\it measured} redshift-space correlation functions $\xi(s,\mu)$
for various cosmological models and quasar survey parameters.
We consider two classes of cosmological models: flat models with
$\om+\omlsp=1$, and open models with $\omlsp=0$.
The $\om=1$, Einstein-de Sitter cosmology is a limiting case of both
families.  We assume that the real-space quasar correlation function
is a power-law, $\xi(r) = (r/r_0)^{-\gam}$, out to $2r_0$.
Once the cosmological model, the correlation function parameters,
and the quasar survey parameters 
are set, the expected number of pairs in a separation/redshift
bin $ds d\mu dz$ is given by equation\r{dN.pairs}.

Because the sparseness of the quasar distribution puts us in the Poisson
limit (see \S 3), we are able to generate realizations of the
measured $\xi(s,\mu)$ very simply, without creating artificial spatial
distributions.  Given the expected number of pairs in a bin from
equation\r{dN.pairs}, the number of pairs in a Monte Carlo realization
is a random deviate drawn from a Poisson distribution with this
mean value.  The pair numbers for each bin can be generated independently.
In practice we use
20 logarithmic bins in $s$, starting at $0.1 s_{0}$ and
going up to $2 s_{0}$.
We use 5 equal bins in $\mu$; because these bins are large, we compute
the predicted number of pairs by numerical integration over the bin.
We work with redshift bins $\Delta z = 0.15$, implying
12 bins for simulations with a 2dF-like $F(z)$ and 22 bins for simulations
with an SDSS-like $F(z)$.
We perform simulations for seven cosmological models: flat and open
low-density models with $\om=0.1, 0.2, 0.4$, and the Einstein-de Sitter
cosmology, $\om=1$.  To fully specify the theoretical model, we must also 
specify the correlation length $s_0$ and the index $\gam$ as functions
of redshift.  Existing observations provide only weak constraints on $\gam$.  
We adopt $\gam=2$, close to the index $\gam=1.8$ of the galaxy
correlation function, and we assume that $\gam$ is independent of redshift.

Even the quasar correlation length $s_0$ is poorly known at present,
because the sparseness of the quasar distribution in current samples
makes measurements of the quasar correlation function very noisy.
The best existing study is probably that of Shanks \& Boyle 
(1994, hereafter SB),
who combine data from three quasar redshift surveys.
For an $\om=1$, $\omlsp=0$ cosmology, 
they find that their correlation function measurements are consistent
with a quasar correlation length $r_0 = 6 {\rm h}_0^{-1}\;{\rm Mpc}$
that remains constant in comoving coordinates, implying
\be
s_0 \equiv \frac{r_0}{g} = 0.002 (1+z)^{3/2},
\lab{s0.om1}
\ee
where we use equation\r{g.def} for $g(z)$.

In order to scale the SB results to other
cosmological models, we note that the property of the correlation
function most robustly constrained by observations is the total
number of correlated pairs.  From equation\r{N.pairs3}, we see
that the number of correlated pairs is
\be
N_{pairs}(z) \propto \frac{r_0^3}{\tilde g \tilde f^2} \propto
\frac{s_0^3 g^2}{f^2} \propto \frac{s_0^3}{h^2},
\lab{Npairs.scaling}
\ee
where we have dropped the factors that are independent of the
spacetime geometry.  For alternative cosmological models,
we therefore wish to hold the combination $s_0^3/h^2$ equal
to the value implied for the $\om=1$, $\omlsp=0$, 
$r_0 = 6{\rm h}_0^{-1}\;{\rm Mpc}$ model that fits the
SB data.  We therefore adopt
\be
s_{0}(z,\om,\oml)=0.002{(1+z)}^{\frac{3}{2}}
{\left[\frac{h(z,\om,\oml)}{h(z,\Omega=1,\oml=0)}\right]}^{\frac{2}{3}}
\lab{s0.general}
\ee
as a fiducial redshift-space correlation length for our simulations.

\subsection{Maximum likelihood determination of parameters}
Given a simulated or observed set of correlation measurements, we are
interested in estimating the true correlation function parameters and
the parameters of the underlying cosmological model.
Suppose that the data consist of pair counts $N_i$ in $i$ bins, where
$i$ may in fact represent a multidimensional (e.g. $s, \mu$) space.
We have a model ${\cal M}$ for the correlation function that may depend
upon several parameters, e.g. ${\cal M} (s_0, \gam, \om, \omlsp)$. 
This model predicts a number of pairs in each bin
$A_i=\bar{n} V_i (1+\xi (s_i, \mu_i)) N_Q$, where $V_i$ is the bin volume.
In the Poisson limit $N_c \la 1$, which is expected to hold for realistic
quasar surveys (see \S 3),
the probability of detecting $N_i$ pairs in bin $i$
when $A_i$ are expected is
\be
P(N_i|A_i)=\frac{e^{-A_i} \cdot A_i^{N_i}}{N_i!}.
\ee
The probabilities for separate bins are independent, so the overall
likelihood ${\cal L}$ of obtaining the data given the model is 
\be
{\cal L} \equiv P(D|{\cal M}) = {\prod}_i \frac{e^{-A_i} 
\cdot A_i^{N_i}}{N_i!},
\lab{likelihood}
\ee
implying
\be
\ln({\cal L}) = {\sum}_i (-A_i + N_i \ln\,A_i- \ln \, N_i!).
\ee
Since the data $N_i$ are independent of the model parameters, one can find
the maximum likelihood model by maximizing the quantity
\be
\ln{\cal L}^{'} ({\cal M})=\sum_i (N_i \ln\,A_i-A_i).
\ee
The relative likelihood of two models ${\cal M}_{1}$ and 
${\cal M}_{2}$ is simply
$\exp(\ln{\cal L}^{'} ({\cal M}_{1})-\ln{\cal L}^{'} ({\cal M}_{2}))$.

Although our focus in this paper is on constraining $\omlsp$, the
maximum-likelihood technique outlined here is quite general.
For a specified cosmology, one can use this technique to estimate
correlation function parameters in a way that makes maximum use
of the available data (e.g., Stephens et al.\ 1997), and the
method can easily be extended to incorporate parametrized descriptions
of peculiar velocity distortions, evolution of clustering, and
so forth.  The sparse sampling limit is crucial to this approach,
for it is only this property that makes it possible to write down
a straightforward expression for the likelihood, equation\r{likelihood}.
For a dense sample like a typical galaxy redshift survey, the likelihood
is a much more complicated function of the model parameters and
the data, and a maximum-likelihood approach is correspondingly more
cumbersome.  However, rich galaxy clusters do provide a sparse tracer 
of structure on large scales, and Croft et al. (1997) have used the 
same maximum-likelihood method to estimate parameters of the 
cluster correlation function.

\subsection{Results}
Figures 5--9 present our main results.  In Figure~5 we examine the
ability of the 2dF (left) and SDSS (right) quasar samples to 
constrain $\om$ in flat, $\om+\omlsp=1$ models.
For each survey and each value of $\om=0.1$, 0.2, 0.4, and 1,
we generated four realizations of measured $\xi(s,\mu)$ by
the Monte Carlo technique described in \S 4.2.
In all cases we set $\gam=2$ and adopt equation\r{s0.general}
for $s_0(z)$, corresponding to $r_0 = 6{\rm h_0}^{-1}\;{\rm Mpc}$
comoving at all $z$ for the $\om=1$, $\omlsp=0$ model and
to $s_0(z=0)=0.002$ in all models.
For each realization, we then determine the best-fit model
parameters by the maximum-likelihood technique described
in \S 4.3, computing likelihoods for a grid of open models and
a grid of flat models with $0 \leq \om \leq 1$ and
varying values of $s_0$ and $\gam$.
Squares show the estimated values of $\om$, denoted
${\hat{\Omega}}_0$, plotted against the true value $\Omega_{true}$
used to generate the realization.  Small offsets are added along
the $\Omega_{true}$ axis to improve clarity.
Vertical line segments delineate the range of $\om$ values
that give likelihood values with 10\% of the maximum likelihood;
this corresponds roughly to a $2\sigma$ confidence interval
for a Gaussian approximation to the likelihood distribution.
In a few cases --- mostly $\om=1$ realizations, which are a limiting
case of both flat and open models --- an open model gave a higher
likelihood than the best-fit flat model.  For these cases
we plot a filled circle at the value of ${\hat{\Omega}}_0$ for the
best-fit open model.


In open cosmologies, the anisotropy of the correlation function is
only weakly sensitive to $\om$ (see Figure~2), so we do not show
the analog of Figure~5 for open models.
Even with flat models the $\om$ constraints are rather loose, but
it is encouraging to see that in the $\om<1$ models, the
$\om=1$ model can be rejected on the basis of quasar clustering
anisotropy alone in nearly every case.

There are a number of methods to constrain $\om$ from the
dynamics of redshift-space galaxy clustering (e.g., Kaiser 1987;
Carlberg et al.\ 1996; Kepner, Summers, \& Strauss 1997).
These methods suffer from a degeneracy between the density parameter
and the ``bias'' of galaxies with respect to mass, but with the
high-precision measurements from the 2dF and Sloan {\it galaxy}
redshift surveys it should be possible to break this degeneracy
and measure $\om$ directly.  We would then like to use quasar
clustering anisotropy to distinguish between open ($\omlsp=0$)
and flat ($\omlsp=1-\om$) cosmologies of known $\om$; dynamical
techniques are insensitive to this distinction because the
cosmological constant represents an unclustered energy component.

Figure~6 presents the prospects for such a test.  For each realization
of each of our low-$\om$ models, we compute the likelihood ratio
$R \equiv {\cal L}_2/{\cal L}_1$, where ${\cal L}_1$ is the likelihood
of the true model (with the correct values of $\om$, $\omlsp$,
$s_0(z)$, and \gam ) and ${\cal L}_2$ is the likelihood of the 
model with the same $\om$, $s_0(z)$, and $\gam$ but the incorrect
geometry.  For $\om=0.1$ or 0.2, the incorrect geometry can be
rejected at $> 10:1$ odds in all of the 2dF realizations and at $\ga 10:1$
odds in the SDSS realizations.  For $\om=0.4$ the difference between
geometries is smaller, and the incorrect model can only be rejected
at $\sim 10:1$ odds or (occasionally) less.  Since the 2dF and SDSS
samples will be independent (with telescopes in the southern and
northern hemispheres, respectively), a pair of $\sim 10:1$ rejections
would constitute a strong statistical result.  However, there are
uncertainties related to peculiar velocity distortions and evolution
of the correlation function that are not incorporated in our current
Monte Carlo experiments (see \S 5 below), so it is not clear that
the geometry distinction will be possible with these samples for
$\om \geq 0.4$.


As equation\r{N.pairs3} demonstrates, the precision with which
$\xi(s,\mu)$ can be measured depends sensitively on the true quasar
correlation length.  While the SB estimate
is based on the most extensive compilation of existing data,
it is nonetheless quite uncertain, and some other studies have
suggested higher values.  Stephens et al. (1997), for example,
find a maximum-likelihood estimate $r_0=20{\rm h}_0^{-1}\;{\rm Mpc}$
(comoving, for $\om=1$, $\lam=0$) for the high-redshift quasars
in the Palomar Transit Grism Survey.  Figures~7 and~8 repeat
the $\om$-constraint and geometry-discrimination tests for 2dF 
(Figure 7) and SDSS (Figure 8) realizations with a correlation 
length double the value implied by equation\r{s0.general}.  This
higher correlation length corresponds to 
$r_0 = 12{\rm h_0}^{-1}\;{\rm Mpc}$
comoving for $\om=1$, $\omlsp=0$, and to $s_0(z=0)=0.004$ for all models.


With the higher correlation length, incorrect geometry models with
$\om=0.4$ are rejected at $\ga 100:1$ odds in all of the 2dF realizations
and all but one of the SDSS realizations.  With $\om \leq 0.2$,
the incorrect geometry models are rejected with formal odds
exceeding $10^6:1$ in all realizations of both surveys.
The $\om$ determinations in flat models are much more precise
than those shown in Figure~5 for the smaller correlation length.
It is of course possible that the true quasar correlation length
is {\it lower} than the SB estimate instead
of higher, in which case the prospects for the clustering anisotropy
method would be worse than those implied by Figures~5 and~6.
The uncertainty in the amplitude of quasar clustering is presently
the main uncertainty in assessing the power of future quasar
surveys to constrain cosmological parameters by this approach.

Figures~5--8 show that the 2dF survey has somewhat greater constraining
power than the SDSS, even though it has 30,000 quasars instead of 80,000.
The high surface density of the 2dF sample is crucial to its performance;
the factor $N_Q^2/A$ in equation\r{N.pairs3} for the number of 
correlated pairs is nearly the same for the 2dF and the SDSS.
The reason the 2dF survey does slightly better is its more
compact redshift distribution $F(z)$.  The SDSS has roughly
equal numbers of quasars per unit redshift from $z=0.4$ to $z=2.2$,
and a significant fraction of the sample lies at $z>2.2$.  For the
2dF survey, a majority of the quasars lie in the range $1 < z < 2.2$.
The number of correlated pairs per unit redshift is proportional
to $[F(z)]^2$, and the more precise measurement of $\xi(s,\mu)$ at
$z \sim 2$ in the 2dF sample wins out over the greater range of
redshifts probed by the SDSS.

As this comparison suggests, the ability of a quasar survey to
measure clustering anisotropy is sensitive to its details,
and there are strategies that can substantially enhance this
ability for a fixed number of quasars.  One possible approach would
be to use multi-color selection techniques to target quasars in
particular redshift ranges, producing peaks in $F(z)$ at specific
redshifts.  Another approach is simply to observe a smaller area
to a fainter magnitude limit, thus increasing $N_Q^2/A$.  Either
strategy improves the survey's sensitivity to geometrical distortion
(at the cost of longer spectroscopic exposures), provided that the
clustering measurements remain in the sparsely sampled regime,
$N_c \la 1$.  Once $N_c$ exceeds one, new quasar pairs contribute
partially redundant information, and there is more to be gained by
expanding the survey's area or redshift range.  (This statement
applies only to the two-point correlation function and its
relatives; clustering measures that are sensitive to higher-order
correlations would continue to benefit from denser sampling.)

As an illustration, we repeat the tests of Figures 5--8 for a hypothetical
high density quasar survey (HDS), now returning to the smaller quasar
correlation length implied by equation\r{s0.general}.
We assume a sample of 30,000 quasars in 200 square degrees --- the
same $N_Q$ as 2dF, but a surface density 3.25 times higher, and $F(z)$
identical to the one of 2dF.
The Sloan southern stripe quasar survey might provide such a sample
if quasars can be identified and their redshifts measured to a 
sufficiently faint limiting magnitude. Figure 9 shows that even with the small 
correlation length, the high-density survey allows strong rejection
of models with incorrect geometry in all realizations with
$\om=0.1$ or 0.2 and in most realizations with $\om=0.4$.


Although we have focused mainly on the sensitivity of planned surveys 
to \om and \oml, we should note that in our experiments the
2dF and SDSS samples constrain $\gamma$ and $s_{0}$ to at least 
10\% accuracy, with typical accuracy of about 5\%. 
In the cases with a high quasar correlation length or a high-density
quasar survey, the maximum-likelihood estimates achieve 2--3\% accuracy 
in recovering $\gamma$ and 1--3\% accuracy in recovering $s_{0}$. 
We can thus expect these future surveys to provide precise measurements
of the basic parameters characterizing the quasar correlation function.

\section{Discussion}
Our results from \S 4.4 show that the ambitious quasar redshift surveys
planned by the 2dF and SDSS teams can make important contributions to
the study of the geometry of spacetime.
In the case of flat cosmological models, some constraints on \om can be 
obtained even for the SB correlation length, and these
constraints become interestingly tight for a high-density quasar survey
or for the 2dF and SDSS if the correlation length is a factor of two
larger than SB's estimate.
Clustering anisotropy does not provide useful constraints on \om in
open models because the distortion parameter $h(z)$ is insensitive
to \om, and there is a near degeneracy in $h(z)$ between an open model with
low \om and a flat model with a higher \om.
The true power of the clustering anisotropy technique comes into play
if \om itself is determined independently by dynamical methods.
Clustering anisotropy then provides a tool for discriminating between
open models with $\omlsp=0$ and flat models with $\omlsp=1-\om$,
a distinction that is difficult to achieve with dynamical methods alone.
We find that clear discrimination between flat and open geometries 
is possible for the 2dF and SDSS samples with the SB correlation length
if $\om \leq 0.2$, but the discrimination is only marginal for $\om=0.4$.
The high-density survey with the SB correlation length or the 2dF or
SDSS for double the SB correlation length provide clear discrimination
for $\om=0.4$.

The results in Figures~5--9 look fairly promising, 
but there are some limitations in our numerical experiments. 
For example, we assumed no evolution in $\gam$ and a fixed form
of $s_0(z)$ (corresponding to no comoving evolution for $\om=1$, $\lam=0$),
so that we could produce one global fit to the clustering and cosmological
parameters using all of the data.
In the real universe, one will have to allow for the possibility of
different evolution of the correlation function parameters.
The maximum-likelihood method described in \S 4.3 adapts easily
to this if the evolution can be described in a parametrized form,
but because there can be some tradeoff between different effects,
adding new parameters will weaken the constraints on \om and \oml.
Also, as we have already noted, our assumed values of $s_{0}$ and \gam, 
while plausible, are highly uncertain. 
If $s_{0}$ and $\gam$ turn out to be larger than we have assumed, then
the statistical power of the clustering anisotropy test will increase; 
if smaller it will decrease.

Probably the most important element missing in our analysis is the
effect of peculiar velocities, since small scale velocity dispersions
and large scale coherent flows can both induce an angular dependence 
in $\xi (s,\mu)$.  To a first approximation, the small-scale dispersion
has the effect of convolving the correlation function along the line of 
sight with the pairwise velocity distribution function 
(Davis \& Peebles 1983; see also Fisher 1995). 
This causes a large distortion of the correlation function for 
galaxies at $z=0$.  However, if $s_0=0.01$ at $z=2$, then the velocity scale 
corresponding to the correlation length is $0.01c = 3000 \,{\rm kms}^{-1}$. 
The dispersion velocity of quasars is unknown, but a plausible value is 
$\sim 300 \,{\rm kms}^{-1}$ --- the velocity dispersion of typical galaxy 
group. The width of the distribution function
is thus very small compared to the correlation length, so for pairs separated
by $\sim s_0$, the distortion due to small scale dispersion should be small.

Coherent large scale motions pose a potentially more serious problem.
As a rough gauge of the importance of those velocities in confusing
geometrical distortion, we can estimate the ratio of $\xi(s,1)/\xi(s,0)$
resulting from geometry and from linear theory peculiar velocities.
For the surveys that we have considered, most of the geometrical information
comes from the redshift $z \sim 2$, so we will take this value as the
fiducial one.  For typical cosmological models, $h(z) \sim 2$ at $z=2$,
which means that $\xi(s,1)/\xi(s,0)\approx 3.5$ for geometrical distortions 
alone (see equation~[\ref{xi.s}]).
In conventional notation, 
\be
\beta \approx \frac{{\Omega}^{0.6}}{b} \lab{beta.def}
\ee
is the linear theory ratio of peculiar velocity convergence 
$-\nabla\cdot {\bf v}_{pec}/H$ to the galaxy density contrast,
where $b$ is the bias parameter relating galaxy and mass fluctuations.
If $\beta \ll 1$,
then equations (10) and (11) of Hamilton (1992) allow us to write
\be
\frac{\xi(s,1)}{\xi(s,0)}=1-\frac{2\beta\gam}{(3-\gam)+2\beta} \lab{xi.ratio}
\ee
We justify equation\r{xi.ratio} in appendix A and argue that the typical
value of $\beta$ (assuming the SB quasar correlation length)
is $\sim 1/6$ in open, flat, and Einstein-de Sitter cosmological models.
With $\gam=1.8$, the ratio\r{xi.ratio} then becomes
\be
\frac{\xi(s,1)}{\xi(s,0)}\approx 0.6
\ee
Thus, the effect of coherent peculiar velocities by no means 
overwhelms the geometric signal, but it is not negligible. 
Analyses of real data will have to fit jointly for peculiar velocity
distortions and geometrical distortions, using their different
angular and redshift dependences to separate them.
The results of Ballinger et al.\ (1996), Matsubara \& Suto (1996),
and Nakamura et al.\ (1997) provide important steps towards this goal,
though it is not clear whether the linear theory formula for velocity
distortions will be adequate near $\xi\sim 1$. 
A suitably parametrized description of peculiar velocity distortions can
easily be incorporated in the maximum-likelihood scheme.
The geometrical distortion is weakest in the models with low values of 
$h$ (low \om, high \omlsp models).
However, these are also the models for which open and flat geometries
are most easily distinguished, so there is good reason to believe that
such distinctions will remain possible even when the effects of peculiar
velocities are taken into account.

There are other routes to measuring or limiting the cosmological constant.
Probably the most solid existing limits come from the statistics of
gravitational lensing (see, e.g., Carroll, Press, \& Turner 1992;
Maoz \& Rix 1993; Kochanek 1996).  These limits are somewhat dependent
on assumptions about the evolution of the galaxy population
(Rix et al.\ 1994), but current results present a strong case that 
$\omlsp < 0.8$.  A promising approach on the horizon is the application of 
the classical magnitude-redshift test to calibrated candles in the form of
high-redshift, Type Ia supernovae (Perlmutter et al. 1997).
This technique is technically challenging, since one must guard carefully
against differential selection biases between low- and and high-redshift
samples, but if these obstacles can be overcome it is one of the
most powerful methods for constraining $\omlsp$.
This approach relies on the
plausible but not incontrovertible assumption that the luminosities
of Type Ia supernovae do not change systematically over the age of
the universe (see von Hippel, Bothun, \& Schommer 1997).
High-resolution maps of the cosmic microwave background, such as those
expected from the MAP and Planck surveyor satellites, offer another
promising route to constraining \omlsp, along with a host of other 
cosmological parameters (see, e.g., Bennett et al. 1995 and 
Jungman et al. 1995), at least in the context of cosmic inflation models.
It is not yet clear which of these approaches will ultimately yield
the best constraints on the cosmological constant, but the 
existence of four essentially independent methods, all of them able
to produce interesting results in the next 5-10 years, is
very encouraging.  Either they will all yield consistent answers,
in which case they provide not just a convincing constraint but
a consistency test of the underlying cosmological picture that motivates
them, or they will not, in which case they point to a breakdown in
the implicit assumptions behind one or more of the methods.

As an extension of this last point, we note that all of our 
2dF and SDSS realizations in Figure~5 are inconsistent at the $>2\sigma$
level with the Euclidean, $\omlsp=1$, $\om=0$ model even for a flat
geometry with $\om$ as low as 0.1 --- and the inconsistency is much
stronger for higher $\om$, a higher quasar correlation length
(Figures~7 and~8), or a higher density quasar survey (Figure 9).
These results imply that the geometrical distortion described in \S 2
should be detected by these surveys, even if the discrimination between
flat and open space geometries is weak.  This distortion is a fundamental
property of the cosmological spacetime curvature predicted by General
Relativity.  If the universe is as we believe it to be, this signature
of curved spacetime should be detected within the next decade, and perhaps
within the next few years.

\acknowledgments

We thank Jim Gunn for seminal discussions at the outset of this project.
We thank Heidi Newberg, Brian Yanny, Don Schneider, and other members
of the SDSS Quasar Working Group for helpful discussions of plans for
the SDSS quasar survey.  We thank Tom Shanks for information about the
2dF quasar survey and for helpful discussions on recent estimates
of quasar clustering.
We acknowledge support from NASA Theory Grant NAG5-3111.

\clearpage

\appendix
\setcounter{equation}{0}
\renewcommand{\theequation}{\thesection\arabic{equation}}
\section{The value of the coherent flow parameter $\beta$}
Hamilton (1992) gives the linear theory distortions of $\xi(s,\mu)$ caused by
peculiar velocities:
\be
\xi(s,\mu)={\xi}_{0}(r)P_{0}(\mu)+{\xi}_{2}(r)P_{2}(\mu)+{\xi}_{4}(r)P_{4}(\mu)
\lab{axi.exp}
\ee
where $P_{i}$ are Legendre polynomials and ${\xi}_{0},{\xi}_{2}$ and 
${\xi}_{4}$ are defined in terms of the growth parameter $\beta$ 
and the volume-averaged correlation function $\bar{\xi}(r)\propto r^{-\gam}$ as
\be
{\xi}_{0}=-\left(1+\frac{2\beta}{3}+\frac{{\beta}^{2}}{5}\right)\frac{\gam-3}{3}\bar{\xi}(r)
\ee
\be
{\xi}_{2}=-\left(\frac{4\beta}{3}+\frac{4{\beta}^{2}}{7}\right)\frac{\gam}{3}\bar{\xi}(r)
\ee
\be
{\xi}_{4}=\frac{8{\beta}^{2}}{35}\frac{\gam (\gam+2)}{3(5-\gam)}\bar{\xi}(r).
\ee
Recall that $\beta$ is defined as
\be
\beta= \frac{{\Omega}^{0.6}}{b}.
\ee
For $\beta\ll 1$, only the first two addends in\r{axi.exp} influence
the result, since  they contain terms proportional to $\beta$ 
but ${\xi}_{4}$ is proportional to ${\beta}^{2}$.
Using the Legendre polynomial definitions $P_{0}=1$ and 
$P_{2}=\frac{3{\mu}^{2}-1}{2}$ and
neglecting all the terms proportional to ${\beta}^2$, we find
\be
\xi(s,1)=\left(1+\frac{2\beta}{3}-\frac{\gam}{3}-\frac{2}{3}\beta\gam \right)\bar{\xi} \lab{xis1}
\ee
\be
\xi(s,0)=\left(1+\frac{2\beta}{3}-\frac{\gam}{3} \right)\bar{\xi}\lab{xis0}
\ee
Dividing\r{xis1} by\r{xis0} yields
\be
\frac{\xi(s,1)}{\xi(s,0)}=1-\frac{2\beta\gam}{(3-\gam)+2\beta}
\ee
equation\r{xi.ratio} of the main text.

For $\om=1$, $s_{0}=0.01$ at $z=2$ corresponds to 6${\mbox{h}^{-1}}$Mpc
comoving, which happens to be roughly the correlation length of bright
galaxies today. Observations suggest that these galaxies have $\beta\sim 0.5$,
which implies a bias parameter $b=2$ for $\om=1$. 
The linear mass fluctuations in an $\om=1$ universe
scale with $z$ as $1/(1+z)$, so the implied bias factor 
for quasars at $z=2$ is $b\sim 2(1+z)\sim 6$.
Thus, 
\be
\beta(\om=1, \oml=0)\sim \frac{1}{6}.
\ee

Now consider an open universe with $\om\sim 0.3$\footnote{We choose $\om=0.3$
to get some idea about both $\om=0.4$ and $\om=0.2$ cases.}. 
The comoving scale $gs_{0}$
corresponding to $s_{0}=0.01$ at $z=2$ is, in this case, larger by a factor
of about 1.5 than in the $\om=1$ model. This implies that quasars are clustered
more strongly than present-day bright galaxies by a factor of about
${1.5}^{1.8} \sim 2$.  In an open universe, the ratio
of the rms present-day mass fluctuation ${\sigma}_{0}$ to the rms
fluctuation $\sigma(z)$ at redshift $z$ is approximately
\be
\frac{{\sigma}_{0}}{\sigma(z)}=1+\frac{2.5\om}{1+1.5\om} \cdot z.
\ee
The bias parameter of galaxies is $b \sim 1$ at $z=0$ 
if $\om\sim 0.3$, so the implied bias parameter for quasars at $z=2$ is
\be
b\sim 2\cdot 1 \cdot \frac{{\sigma}_{0}}{\sigma(z)}= 4 .
\ee
At $z=2$, the value of the density parameter $\Omega$ is
\be
\Omega=\frac{(1+z)\om}{1+\om z}\approx 0.56.
\ee
Putting these results together, we get $\beta$ for quasars at $z=2$
in a low-density, open universe:
\be
\beta(\om=0.3, \oml=0)\sim \frac{0.71}{4}\approx 0.18.
\ee

The analysis for a low-density flat universe with $\om\sim 0.3$ follows
similar lines.  The correlation length is larger by a factor of about 
$1.8$ relative to the $\om=1$ model. The ratio of mass fluctuations 
at $z=2$ to fluctuations at $z=0$, computed by the appropriate numerical
integral, is
\be
\frac{{\sigma}_{0}}{\sigma(z)}=2.38.
\ee
The implied bias parameter for quasars at $z=2$ is
\be
b\sim {1.8}^{1.8} \cdot 2.38 \approx 6.86.
\ee
At $z=2$, the value of $\Omega$ is $0.92$, and consequently
\be 
\beta(\om=0.3, \oml=0.7) \sim 0.14.
\ee

In all three cases we find 
\be
\beta(\om,\oml,z=2)\sim \frac{1}{6},
\ee
so the assumption that $\beta\leq 0.2$ for quasars at $z=2$ 
is fairly secure, unless the quasar correlation length is significantly 
{\it smaller} than the SB value (in which case detection of 
geometrical distortions of the correlation function may be rather difficult).

\clearpage

\figcaption[fig1.ps]{Geometry of an observer and a quasar pair in redshift 
space, with the distances $s$, $s_{\theta}$ and $s_{z}$ indicated.
\label{fig1}}

\figcaption[fig2.ps]{The distortion parameter $h(z)$ as a function 
of redshift, for the indicated combinations of
density parameter \om and cosmological constant contribution \oml.
A cosmological model with $\Omega_0=0$, $\lambda_0=1$ would
have $h(z)=1$ at all redshifts. \label{fig2}}

\figcaption[fig3.ps]{The angular dependence of the correlation function 
for different values of the distortion parameter $h$. \label{fig3}}

\figcaption[fig4.ps]{F(z) for 2dF (left panel) and SDSS (right panel)
with superimposed dashed curves that correspond to models used
in our simulations. \label{fig4}}

\figcaption[fig5.ps]{Constraints on $\om$ for flat ($\om+\omlsp=1$) 
cosmological models in simulations of the 2dF (left) and Sloan (right) quasar 
samples. Eight realizations (four for each survey) were generated for each
value of $\om=0.1$, 0.2, 0.4, and 1.  In all cases the true
correlation function was characterized by $\gam=2$, $s_0(z=0)=0.002$.
Open squares show the maximum-likelihood estimates of $\om$; error
bars delineate the range of models with likelihood greater than 10\%
of the maximum likelihood.  Filled circles are plotted in cases
where an open model had a higher likelihood than the best-fit flat
model.  Small horizontal offsets are introduced for clarity. \label{fig5}}

\figcaption[fig6.ps]{Relative likelihoods of flat and open models for 
simulations of the 2dF (left) and Sloan (right) quasar samples.  Triangles 
represent flat models; $R^{-1}$ is the ratio of the likelihood of the true
(flat geometry) model to the likelihood of an open model with the
same value of $\om$ and correlation function parameters.
Circles show the corresponding comparison for intrinsically open models.
For $\om=0.1$ or 0.2, the incorrect geometry models can usually be
rejected with 10:1 confidence or better.  Small horizontal offsets 
have been introduced for clarity. \label{fig6}}

\figcaption[fig7.ps]{Constraints from simulations of the 2dF quasar sample 
for a high quasar correlation length, 
$s_0(z=0)=0.004$ ($r_0(z=0)=12 {\rm h}_0^{-1}\;{\rm Mpc}$).
Left panel shows the ratio of the likelihood of a model with incorrect
geometry to the true model, in the same format as Figure~6.
Right panel shows constraints on $\om$ for flat models, in the
same format as Fig.~5. \label{fig7}}

\figcaption[fig8.ps]{Same as Fig. 7, but for simulations of the Sloan sample.
\label{fig8}}

\figcaption[fig9.ps]{Same as Figs. 7 and 8, but for 
$s_0(z=0)=0.002$ ($r_0(z=0)=6 {\rm h}_0^{-1}\;{\rm Mpc}$)
and sample parameters of a hypothetical high-surface-density
quasar survey, with 30,000 quasars in 200 square degrees.
\label{fig9}}


\clearpage

\begin{references}
\reference{alc} Alcock, C., Paczy\'{n}ski, B. 1979, Nature, 281, 358
\reference{bal} Ballinger, W.E., Peacock, J.A., Heavens, A.F., astro-ph/9605017
\reference{ben} Bennett, C.L., Hinshaw, G., Jarosik, N., Mather, J.S., Meyer, S.S., Page, L., Skillman, D., Spergel, D.N., Wilkinson, D.T., Wright, E.L. 1995, BAAS, 187, 7109B
\reference{boy1} Boyle, B. J. 1991, in Ann. NY Acad. Sci. 647, 
Texas/ESO-CERN Symposium on Relativistic Astrophysics, Cosmology, 
and Fundamental Physics, ed. J. D. Barrow, L. Mestel, \& P. A. Thomas, 14
\reference{boy2} Boyle, B. J., Shanks, T., Peterson, B. A. 1988, \mnras, 235, 935
\reference{carl} Carlberg, R. G., Yee, H. K. C., Ellingson, E.,
Abraham, R., Gravel, P., Morris, S., \& Pritchet, C. J. 1996, \apj, 462, 32
\reference{carr} Carroll, S. M., Press, W. H., \& Turner, E. L. 1992, \araa, 30, 499
\reference{cro} Croft, R.A.C., Dalton, G.B., Efstathiou, G., Sutherland, W., 
Maddox, S., \mnras, submitted, astro-ph/9701040
\reference{dav} Davis, M., Peebles, P.J.E., 1983, \apj, 267, 465
\reference{fis} Fisher, K. B. 1995, \apj, 448, 494
\reference{ham} Hamilton, A.J.S., 1992, \apj, 385, L5
\reference{jun} Jungman, G., Kamionkowski, M., Kosowsky, A., \& Spergel, D. N. 1995,
Phys Rev D, 54, 1332
\reference{kai} Kaiser, N. 1987, \mnras, 227, 1
\reference{kep} Kepner, J., Summers, F., \& Strauss, M. 1997, \apj, in press,
astro-ph/9607097
\reference{koc} Kochanek, C. 1996, \apj, 466, 638
\reference{mao} Maoz, D., \& Rix, H. 1993, \apj, 416, 425
\reference{mat} Matsubara, T., Suto, Y., 1996, \apj, 470L, 1M
\reference{nak} Nakamura, T. T., Matsubara, T., \& Suto, Y. 1997, \apj, submitted,
astro-ph/9706034
\reference{pee} Peebles, P.J.E. 1980, The Large Scale Structure of the Universe,
	   (Princeton: Princeton University Press)
\reference{per} Perlmutter, S. et al. 1997, \apj, 483, 565
\reference{phi} Phillipps, S. 1994, \mnras, 269, 1077
\reference{rix} Rix, H., Maoz, D., Turner, E.L., \& Fukugita, M. 1994, \apj, 435, 49
\reference{ryd1} Ryden, B.S. 1995a, \apj, 452, 25R
\reference{ryd2} Ryden, B.S., Melott, A., 1996, \apj, 470, 160R
\reference{sha} Shanks, T., Boyle, B.J., 1994, \mnras, 271, 763 (SB)
\reference{son} Soneira, R. M., Peebles, P.J.E., 1978, AJ, 83, 845
\reference{ste} Stephens, A.W., Schneider, D.P., Schmidt, M., Gunn, J.E., 
Weinberg, D.H., 1997, AJ, in press
\reference{von} von Hippel, T., Bothun, G. D., \& Schommer, R. A., \aj, in press,
astro-ph/9706113
\reference{war} Warren, S. J., Hewett, P. C., Osmer, P. S. 1994, \apj, 421, 412
\reference{wei} Weinberg, S. 1972, in ``Gravitation and Cosmology'' (New York: Wiley)
\end{references}
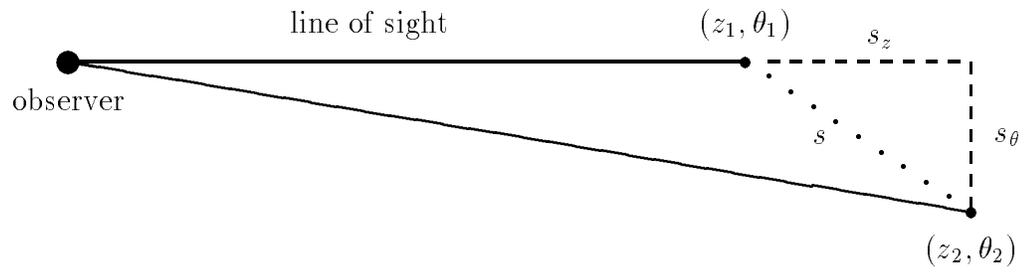
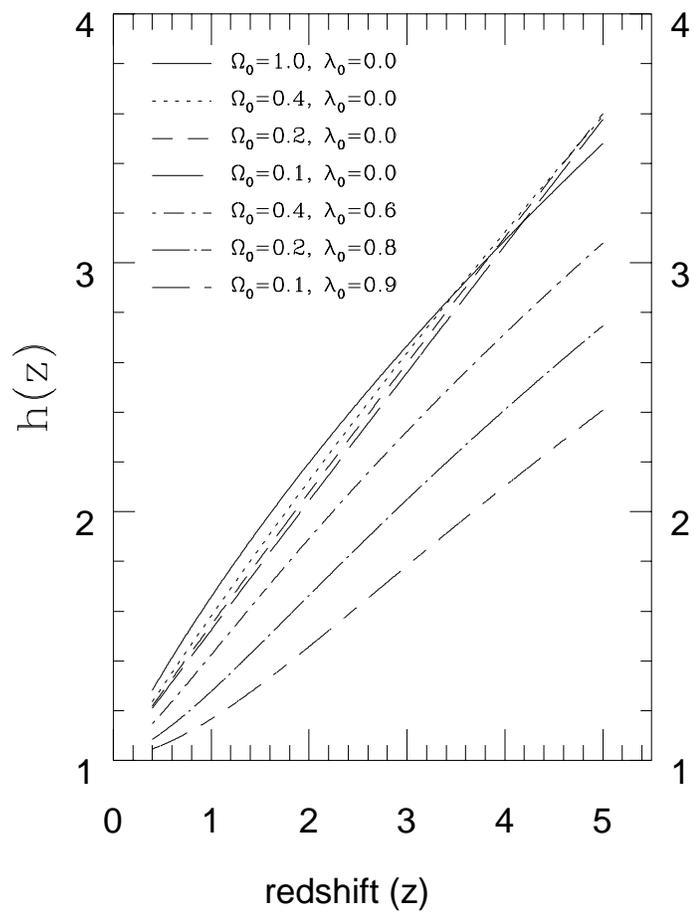
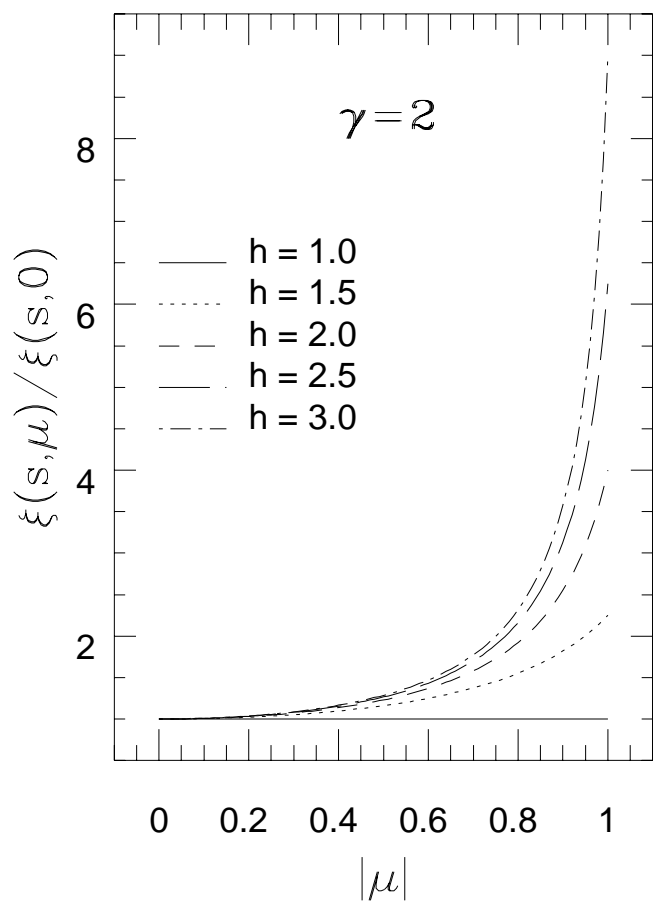
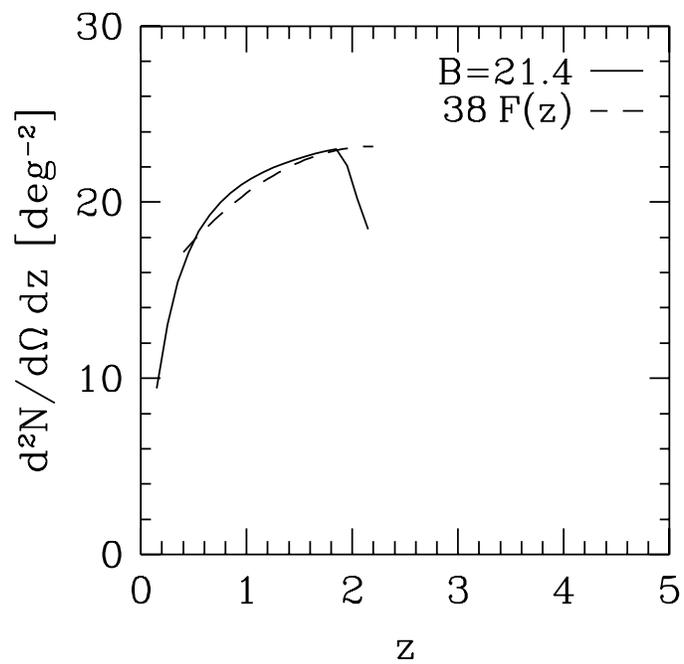
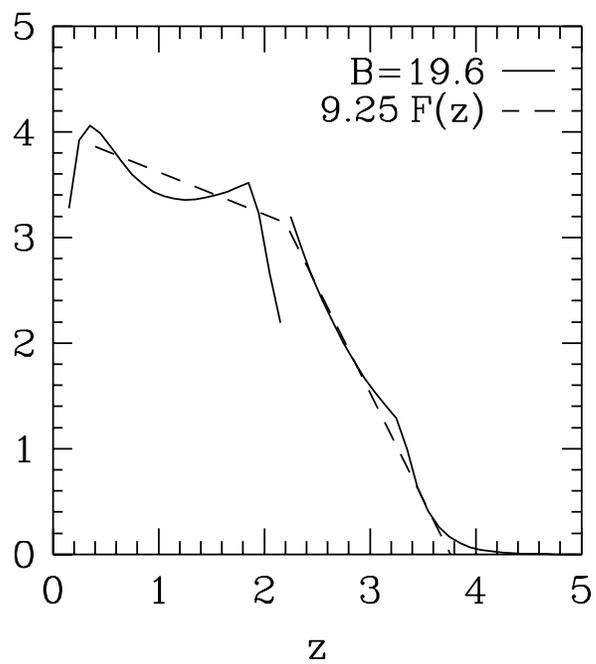
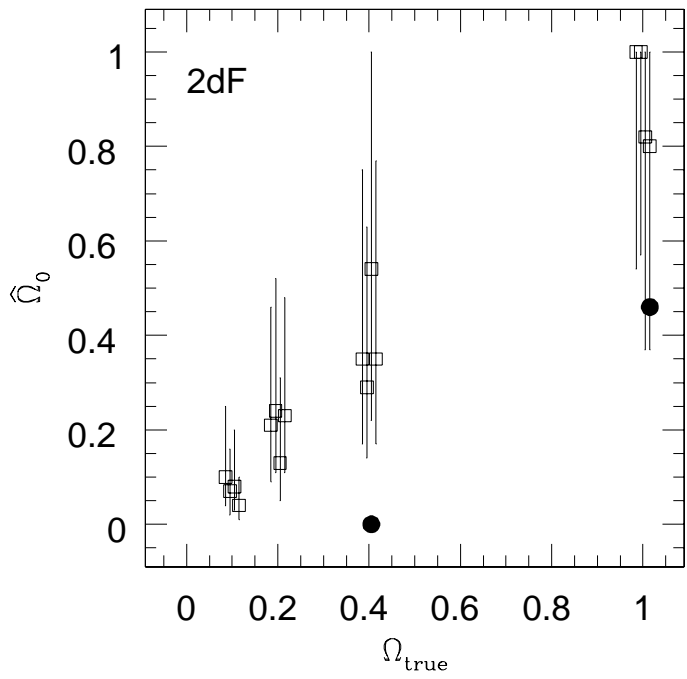
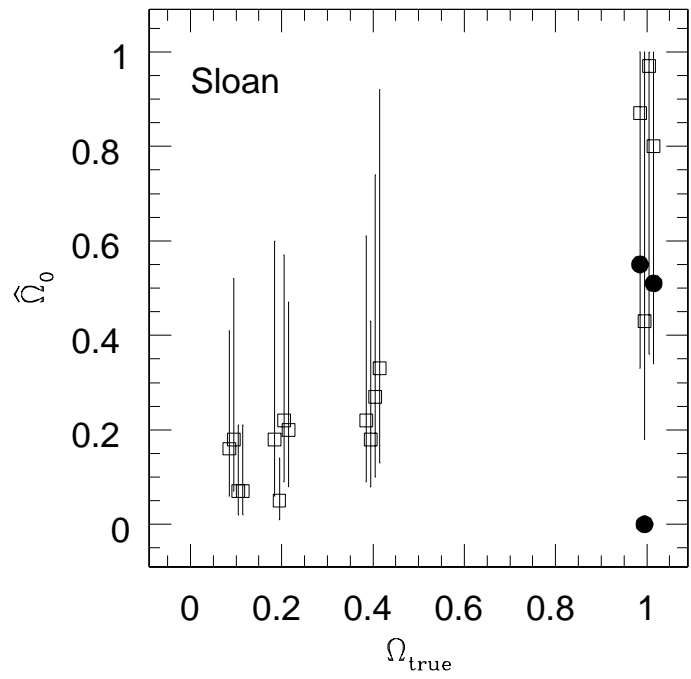
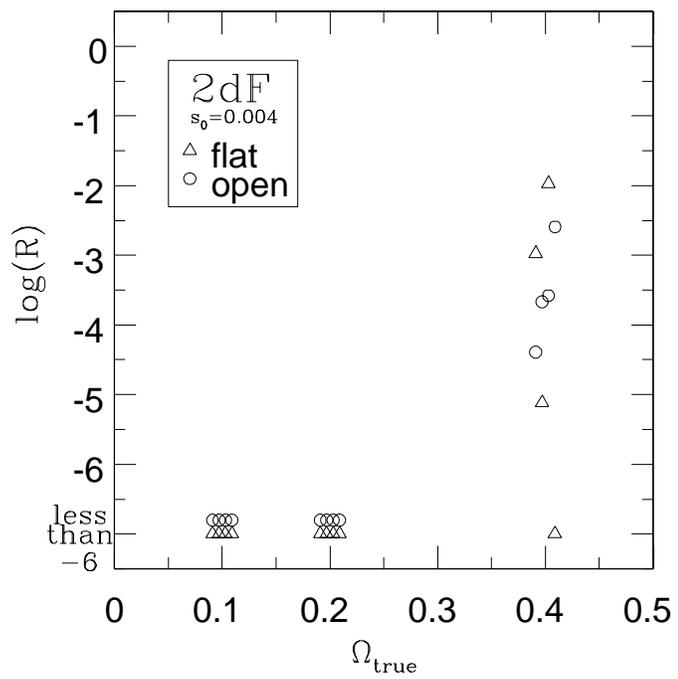
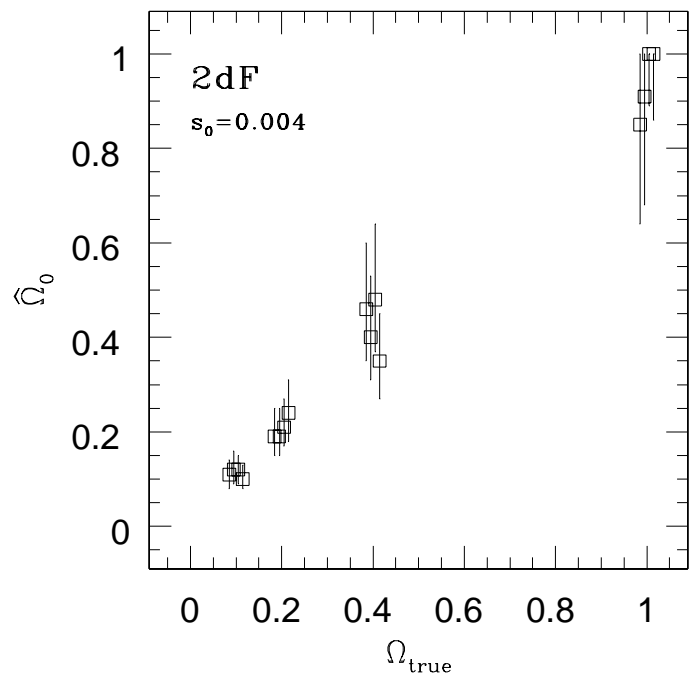
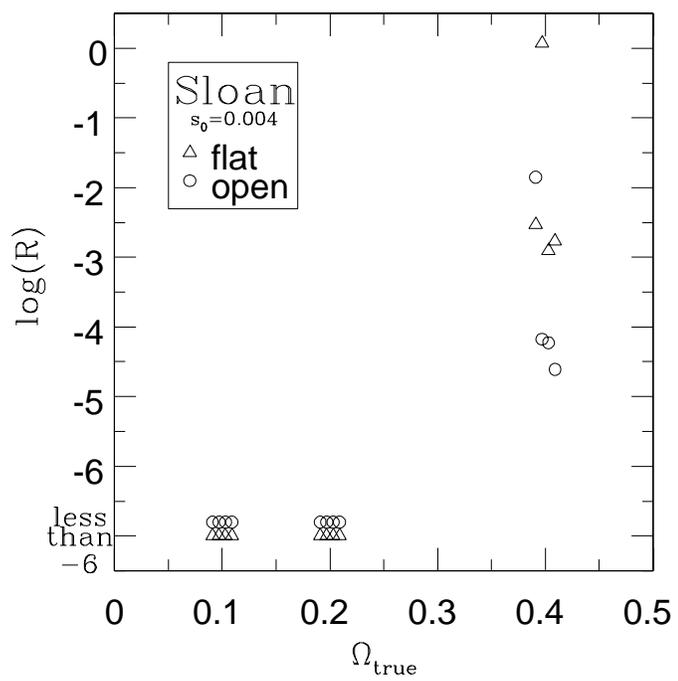
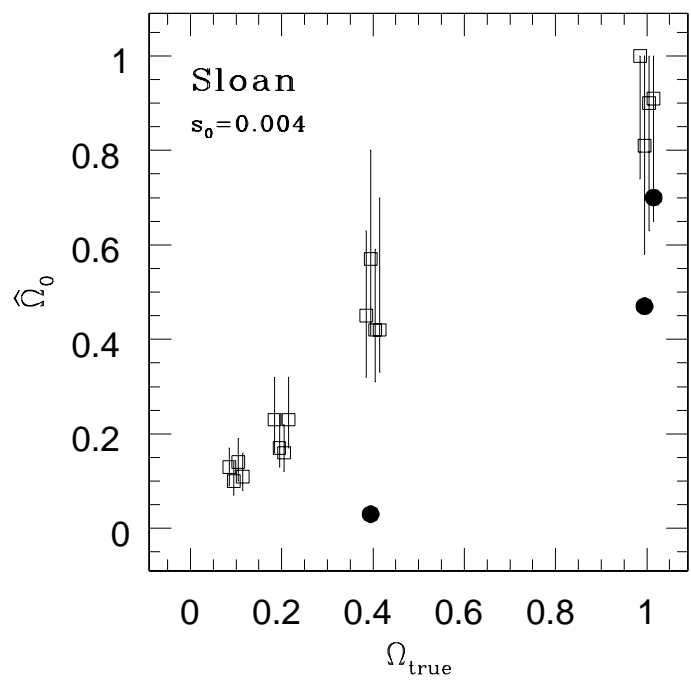
\end{document}